\documentclass[twocolumn]{aastex63}

\received{June 1, 2099}
\revised{January 10, 2099}
\accepted{\today}
\submitjournal{ApJ}

\shorttitle{X-rays from the PeVatron candidate}
\shortauthors{Fujita et al.}

 \usepackage[T1]{fontenc}

\begin{document}

\title{X-Ray Emission from the PeVatron-candidate Supernova Remnant
G106.3+2.7}

\correspondingauthor{Yutaka Fujita}
\email{y-fujita@tmu.ac.jp}

\author[0000-0003-0058-9719]{Yutaka Fujita}
\affiliation{Department of Physics, Graduate School of Science,
Tokyo Metropolitan University, 1-1 Minami-Osawa,\\
Hachioji-shi, Tokyo 192-0397, Japan}

\author{Aya Bamba} \affiliation{Department of Physics, Graduate School
of Science, The University of Tokyo, 7-3-1 Hongo, Bunkyo-ku, Tokyo
113-0033, Japan} 
\affiliation{Research Center for the Early Universe,
School of Science, The University of Tokyo, 7-3-1 Hongo, Bunkyo-ku,
Tokyo 113-0033, Japan }

\author[0000-0002-0726-7862]{Kumiko K. Nobukawa} \affiliation{Faculty of Science and
Engineering, Kindai University, 3-4-1 Kowakae, Higashi-Osaka,
577-8502, Japan}

\author{Hironori Matsumoto} \affiliation{Department of Earth and Space
Science, Graduate School of Science, Osaka University,\\ 1-1
Machikaneyama-cho, Toyonaka, Osaka 560-0043, Japan}
\affiliation{Project
Research Center for Fundamental Sciences, Graduate School of Science,
Osaka University,\\ 1-1 Machikaneyama-cho, Toyonaka, Osaka 560-0043,
Japan}


\begin{abstract}

We report the discovery of diffuse X-ray emission around the supernova
remnant (SNR) G106.3+2.7, which is associated with VER~J2227+608 and
HAWC~J2227+610 and is known as a candidate for a PeV cosmic-ray
accelerator (PeVatron). We analyze observational data of Suzaku around
the SNR and the adjacent pulsar PSR J2229+6114. We find diffuse X-ray
emission that is represented by either thermal or nonthermal
processes. However, the metal abundance for the thermal emission is
$<0.13\: Z_\sun$, which may be too small in the Milky Way and suggests
that the emission is nonthermal. The intensity of the diffuse emission
increases toward PSR J2229+6114 in the same way as radio emission, and
it is in contrast with gamma-ray emission concentrated on a molecular
cloud. The X-ray photon index does not change with the distance from the
pulsar and it indicates that radiative cooling is ineffective and
particle diffusion is not extremely slow. The X-ray and radio emissions
seem to be of leptonic origin and the parent electrons may originate
from the pulsar. The gamma-ray emission appears to be of hadronic origin
because of its spatial distribution. The parent protons may be tightly
confined in the cloud separately from the diffusing electrons.

\end{abstract}


\section{Introduction} 
\label{sec:intro}

The origins of cosmic rays (CRs) are still not fully understood. It is
often believed that CRs with energies up to a few PeV (so-called
“knee”) originate from sources within the Milky Way.  Although it is
generally accepted that CRs in the GeV and TeV energy ranges are
accelerated by shocks in supernova remnants (SNRs), no SNR has been
shown to emit gamma rays to hundreds of TeV as would be indicative of a
source capable of accelerating protons to PeV (PeVatron).

The SNR G106.3+2.7 may be a possible exception. It was discovered by a
408 MHz radio survey \citep{1990A&AS...82..113J}, and then it was
interpreted as an SNR with an unusual head-tail morphology
\citep{2000AJ....120.3218P}. The diffuse radio emission is bright at the
``head'' that is close to the powerful pulsar PSR~J2229+6114
(Figure~\ref{fig:map}) with a spin-down luminosity of $L_{\rm sp} =
2.2\times 10^{37}\rm erg~s^{-1}$ \citep{2001ApJ...552L.125H}. TeV
gamma-ray emission from this region has been discovered by the Milagro
collaboration at 20~TeV \citep{2007ApJ...664L..91A} and 35~TeV
\citep{2009ApJ...700L.127A}, and by the VERITAS collaboration at $\sim
1$--15~TeV \citep{2009ApJ...703L...6A}. The VERITAS source was named as
VER~J2227+608.  Recently, HAWC identified the object as HAWC~J2227+610
at least up to $\sim 100$~TeV \citep{2020ApJ...896L..29A}, which means
that the energy of the parent protons should be close up to
$\sim$~PeV. It has also been detected in the GeV band with Fermi
\citep{2019ApJ...885..162X}. These gamma-ray observations have shown
that the emission is concentrated around a molecular cloud, which
suggests that the gamma rays are produced through the interaction
between CR protons and protons in the molecular cloud ($\pi^0$ decay or
hadronic origin). The gamma-ray emitting region is located in the
``tail'' of the diffuse radio emission (Figure~\ref{fig:map}). The spin
axis of PSR~J2229+6114 is toward the northwest in contrast with the
direction of the tail (southwest; \citealt{2006ApJ...638..225K}).

X-ray observations can be used to discuss whether the gamma rays are
produced by CR electrons (leptonic origin).  Moreover, they are useful
to constrain the maximum energy of CR electrons if the X-ray emission is
synchrotron \citep{2019ApJ...885..162X,2020ApJ...897L..34L}. However,
previous X-ray observations were limited to the emissions from
PSR~J2229+6114 and its vicinity \citep{2001ApJ...547..323H}. In this
paper, we report the results of Suzaku observations of
G106.3+2.7. The Suzaku X-ray Imaging Spectrometer (XIS)
\citep{2007PASJ...59S..23K} has high sensitivity and stable low
background \citep{2006SPIE.6266E..42Y}, and it is the optimal device to
search for diffuse X-ray emission from G106.3+2.7. 

This paper is organized as follows. In Section~\ref{sec:obs}, we
summarize the Suzaku observations. The data analysis routines are
presented in Section \ref{sec:ana} and the results are shown in
Section~\ref{sec:result}. The origin of the nonthermal emissions is
discussed in Section~\ref{sec:disc}, followed by a summary in
Section~\ref{sec:sum}. The errors quoted in the text and tables all
represent a $1\:\sigma$ confidence level.

\section{Observations}
\label{sec:obs}

\begin{figure}[t]
\plotone{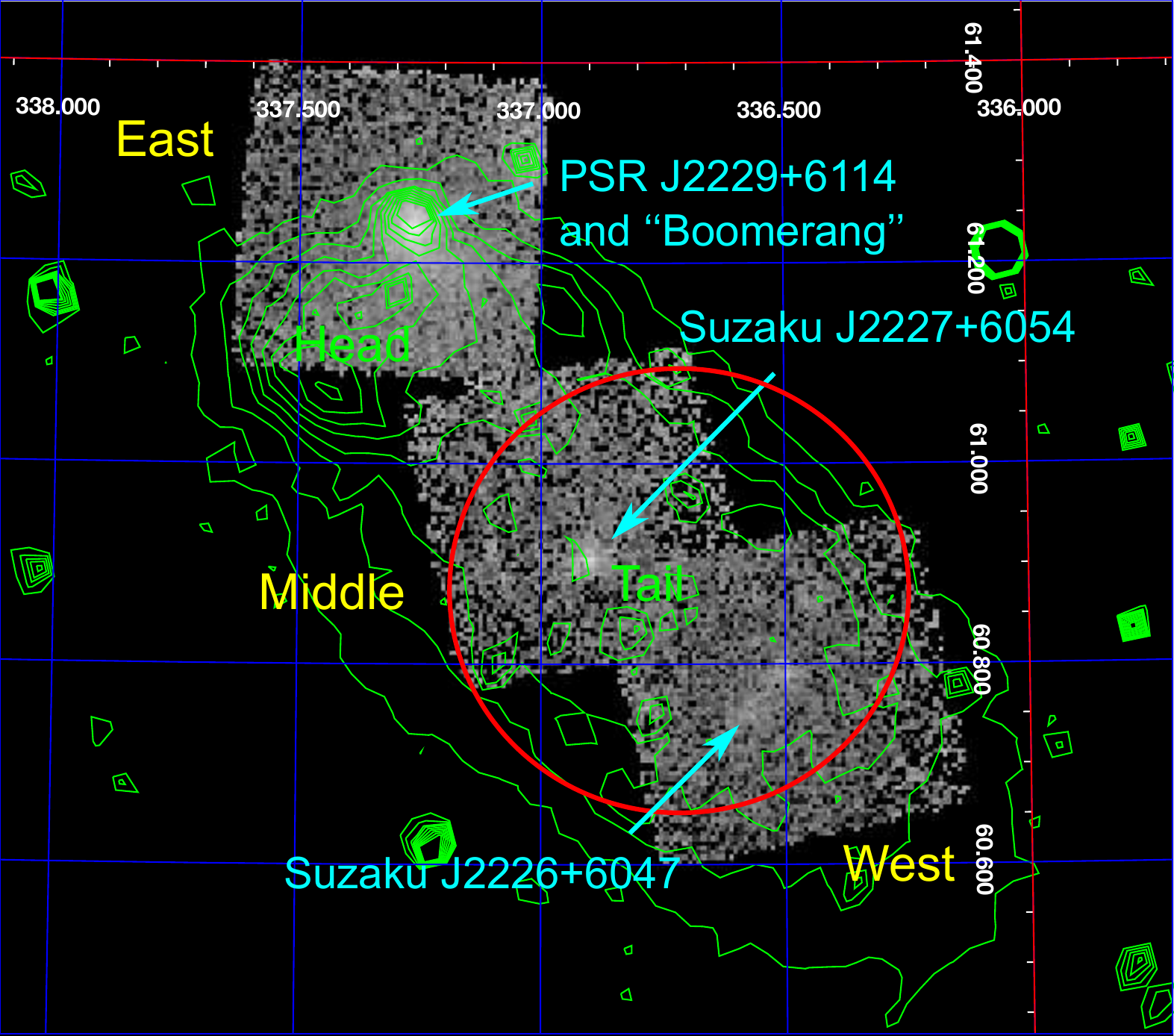} \caption{Combined XIS~1+3 image of G106.3+2.7 in the
1--5~keV band. The brightness scale is logarithmic. Vignetting
correction was applied after subtracting non-X-ray backgrounds. The red
circle approximately represents the extended gamma-ray emitting region
\citep{2009ApJ...703L...6A,2019ApJ...885..162X}. Green contours
represent the radio continuum emission at 1420 MHz by the Canadian
Galactic Plane Survey (CGPS; \citealt{2003AJ....125.3145T}). Contour
levels are 6.6, 6.8, 7.0, 7.2, 7.5, 7.8, 8.1, 8.4, and 8.7~K.
\label{fig:map}}
\end{figure}

We analyzed the archive data of Suzaku XIS that cover three fields
around G106.3+2.7. They are called the ``East,'' ``Middle,'' and
``West,'' respectively (Figure~\ref{fig:map}). Since the possible
diffuse X-ray source could cover the entire Suzaku fields, we also use
the Suzaku data of a blank field near G106.3+2.7 that is called the
``Reference.'' From now on, we collectively refer to the East, Middle,
and West fields as the ``Source'' fields. The observation logs are
summarized in Table~\ref{tab:log}.  The XIS is composed of four CCD
cameras (XIS~0, 1, 2, and 3). XIS~0, 2, and 3 are front-illuminated (FI)
CCDs, and XIS~1 is a back-illuminated (BI) CCD. The field of view (FOV)
of the XIS is $17\farcm 8\times 17\farcm 8$. The entire FOV of XIS~2 and
one-fourth of XIS~0 havej been out of function since 2006 November and
2009 June, respectively. We chose cleaned event data that had been
screened in standard ways. We used HEASoft~6.25
\citep{2014ascl.soft08004N} for data analysis.

Figure~\ref{fig:map} is a mosaic XIS image of G106.3+2.7 in the 1--5~keV
band. Vignetting is corrected. The XIS fields fairly cover the TeV
gamma-ray emitting region
\citep{2009ApJ...703L...6A,2019ApJ...885..162X}. The bright source
around the center of the East field is the pulsar PSR~J2229+6114. The
Middle and West fields also contain X-ray sources near their centers,
respectively. We have checked Chandra and XMM-Newton images
and have found that the source in the Middle field (Suzaku J2227+6054)
is a compact source. Although that in the West field (Suzaku J2226+6047)
appears to be diffuse, it contains a point source.

\section{Spectral Analysis}
\label{sec:ana}

In the spectral analysis, we exclude PSR~J2229+6114 and its pulsar wind nebula (PWN) by a $260''$ radius
and remove Suzaku J2226+6047 and Suzaku J2227+6054 by a $180''$ radius
because we are interested in diffuse emission. We also exclude other
point sources with fluxes of $>2\times 10^{-14}\rm\: erg\: cm^2\:
s^{-1}$ listed in Chandra Source Catalog 2.0\footnote{https://cxc.harvard.edu/csc/} and XMM-DR9 catalog\footnote{http://xmmssc.irap.omp.eu/Catalogue/4XMM-DR9/4XMM\_DR9.html} by a
$90''$ radius. We analyze photons with energies of 0.5--7~keV in the
entire fields excluding those point sources.

\begin{deluxetable*}{ccccc}
\label{tab:log}
\tablecaption{Observation Log}
\tablewidth{0pt}
\tablehead{
\colhead{Name} & \colhead{ObsID} & \colhead{Date} & \colhead{Position} & \colhead{Exposure}  \\
\colhead{} & \colhead{} & \colhead{(YYYY MM DD)} & \colhead{(J2000)} & \colhead{(ks)} 
}
\startdata
East  & 505054010 & 2010 05 16 & (337.33, 61.23) & 59 \\
Middle& 505072010 & 2010 08 15 & (336.95, 60.94) & 25 \\
West  & 505073010 & 2010 08 16 & (336.54, 60.77) & 55 \\
Reference & 501100010 & 2006 06 06 & (347.95, 61.94) & 72 \\
\enddata
\end{deluxetable*}

\begin{deluxetable*}{cccccccc}
\label{tab:fit}
\tablecaption{Fitting Results}
\tablewidth{0pt}
\tablehead{
\colhead{Field (Emission)} & \colhead{$kT_{\rm GHG}^a$} & \colhead{$N_{\rm H,c}^b$} & \colhead{$\Gamma^c$} & \colhead{$kT^d$} & \colhead{$Z^e$} & \colhead{$\langle I\rangle^f$} & \colhead{$\chi^2$/dof$^g$} \\
 & \colhead{(keV)} & \colhead{($10^{22}\rm\: cm^{-2}$)} & \colhead{(keV)} & \colhead{($Z_\odot$)} & &  &  
}
\startdata
East (Nonthermal) & $0.67_{-0.03}^{+0.03}$ & $0.88_{-0.07}^{+0.06}$ & $2.2_{-0.1}^{+0.1}$ & \nodata & \nodata & $5.4_{-0.2}^{+0.2}$ & 552.98/433 \\
East (Thermal) & $0.67_{-0.04}^{+0.03}$ & $0.71_{-0.05}^{+0.05}$ & \nodata & $3.8_{-0.4}^{+0.4}$ & $0.12_{-0.11}^{+0.11}$ & $4.8_{-0.3}^{+0.4}$ & 546.20/433 \\
Middle (Nonthermal) & $0.70_{-0.05}^{+0.03}$ & $0.89_{-0.12}^{+0.13}$ & $2.3_{-0.2}^{+0.2}$ & \nodata & \nodata & $3.4_{-0.3}^{+0.3}$ & 348.58/241 \\
Middle (Thermal) & $0.70_{-0.05}^{+0.02}$ & $0.69_{-0.08}^{+0.09}$ & \nodata & $3.5_{-0.7}^{+0.8}$ & $0.00_{-0.00}^{+0.16}$ & $3.0_{-0.4}^{+0.6}$ & 348.46/241 \\
West (Nonthermal) & $0.65_{-0.03}^{+0.06}$ & $0.90_{-0.11}^{+0.12}$ & $2.2_{-0.2}^{+0.2}$ & \nodata & \nodata & $2.4_{-0.2}^{+0.2}$ & 486.01/386 \\
West (Thermal) & $0.65_{-0.03}^{+0.05}$ & $0.75_{-0.07}^{+0.07}$ & \nodata & $3.9_{-0.6}^{+1.1}$ & $0.00_{-0.00}^{+0.12}$ & $2.1_{-0.3}^{+0.4}$ & 486.32/386 \\
Combined (Nonthermal) & $0.64_{-0.02}^{+0.03}$ & $0.89_{-0.06}^{+0.05}$ & $2.2_{-0.1}^{+0.1}$ & \nodata & \nodata & $3.7_{-0.1}^{+0.1}$ & 936.64/834 \\
Combined (Thermal) & $0.64_{-0.03}^{+0.03}$ & $0.71_{-0.03}^{+0.04}$ & \nodata & $3.7_{-0.03}^{+0.03}$ & $0.05_{-0.05}^{+0.08}$ & $3.3_{-0.4}^{+0.4}$ & 931.24/834 \\
\enddata
\tablecomments{$^a$ Temperature of the diffuse Galactic hot gas (GHG). $^b$
Combined absorption column density by the interstellar medium and
possible dense gas around G106.3+2.7. $^c$ Photon index of the
nonthermal emission from G106.3+2.7. $^d$ Temperature of the thermal
emission from G106.3+2.7. $^e$ Metal abundance of thermal emission from
G106.3+2.7. $^f$ Average surface brightness of the source component in
the 2--10~keV band ($10^{-15}~\rm erg~cm^{-2}~s^{-1}~arcmin^{-2}$). $^g$
$\chi^2$ and the degree of freedom for the spectral fit.}
\end{deluxetable*}

The response and auxiliary files were created using {\tt xisrmfgen} and
{\tt xissimarfgen} \citep{2007PASJ...59S.113I} in the HEADAS package,
assuming that the emission uniformly comes from the individual
fields. The spectra of both the Source and Reference fields contain the
non-X-ray background (NXB) and cosmic X-ray background (CXB).  Other
background components such as the diffuse Galactic hot gas (GHG)
including the local hot bubble \citep{2009PASJ...61..805Y} may also need
to be considered at $\lesssim 1$~keV. The NXB is generated by {\tt
xisnxbgen} \citep{2008PASJ...60S..11T} and it is applied for both the
Source and Reference fields.  We assume that H{\footnotesize I} column
density due to the interstellar absorption is $N_{\rm H,i}=0.8\times
10^{22}\rm\: cm^{-2}$ \citep{2016A&A...594A.116H}; the slight variation
among the fields ($< 0.1\times 10^{22}\rm\: cm^{-2}$) is ignored. For
the intrinsic emission from the Source fields, we consider nonthermal or
thermal processes. In the former case, the spectrum of the Source fields
after subtracting NXB is given by XSPEC \citep{1996ASPC..101...17A}
models:
\begin{eqnarray}
\label{eq:specNTH}
S_{\rm NTH} & =& {\tt phabs}\langle 1\rangle
*{\tt apec}\langle 2\rangle  \nonumber \\
& +&  {\tt phabs}\langle 3\rangle ({\tt
pegpwrlw}\langle 4\rangle + {\tt pegpwrlw}\langle 5\rangle).
\end{eqnarray}
where {\tt phabs}$\langle 1\rangle$ is the interstellar absorption
($N_{\rm H,i}=0.8\times 10^{22}\rm\: cm^{-2}$), {\tt apec}$\langle
2\rangle$ represents the GHG with the solar metal abundance, {\tt
phabs}$\langle 3\rangle$ is the combined absorption by the interstellar
medium and possible dense gas around G106.3+2.7 ($N_{\rm H,c}$), {\tt
pegpwrlw}$\langle 4\rangle$ corresponds to the CXB given by a power law,
and {\tt pegpwrlw}$\langle 5\rangle$ is the nonthermal emission from
G106.3+2.7. We represent the CXB spectrum (${\tt pegpwrlw}\langle
4\rangle$) with a power-law model with a fixed photon index of 1.4 and
set the normalization free because the intensity of CXB can vary from
field to field \citep{2002PASJ...54..327K}. The photon index of {\tt
pegpwrlw}$\langle 5\rangle$ is given by $\Gamma$.

If we instead consider thermal emission from
G106.3+2.7, {\tt pegpwrlw}$\langle 5\rangle$ in
equation~(\ref{eq:specNTH}) is replaced by {\tt apec}$\langle 5\rangle$,
and the spectrum is written as
\begin{eqnarray}
\label{eq:specTH}
S_{\rm TH} & =& {\tt phabs}\langle 1\rangle
*{\tt apec}\langle 2\rangle  \nonumber \\
& +&  {\tt phabs}\langle 3\rangle ({\tt
pegpwrlw}\langle 4\rangle + {\tt apec}\langle 5\rangle).
\end{eqnarray}
The temperature and the metal abundance of {\tt apec}$\langle 5\rangle$
are given by $kT$ and $Z$, respectively.

We simultaneously fit the spectra of the Source fields and that of the
Reference field. The spectral parameters of the Reference field,
including the normalizations of ${\tt apec}\langle 2\rangle$ and ${\tt
pegpwrlw}\langle 4\rangle$, are constrained to be the same as those of
the Source fields\footnote{The obtained parameters for the Reference
field slightly differ among the fits for the three Source fields
(e.g. $kT_{\rm GHG}$ in Table~\ref{tab:fit}).} except that the fifth
component is zero ({\tt pegpwrlw}$\langle 5\rangle=0$ or {\tt
apec}$\langle 5\rangle=0$) and ${\tt phabs}\langle 3\rangle={\tt
phabs}\langle 1\rangle = N_{\rm H,i}$. We fit the spectra of XIS~0, 1,
and 3 together with the same parameters. We use solar abundances from
\citet{2009LanB...4B..712L}.

\section{Results}
\label{sec:result}

\begin{figure}[t]
\plotone{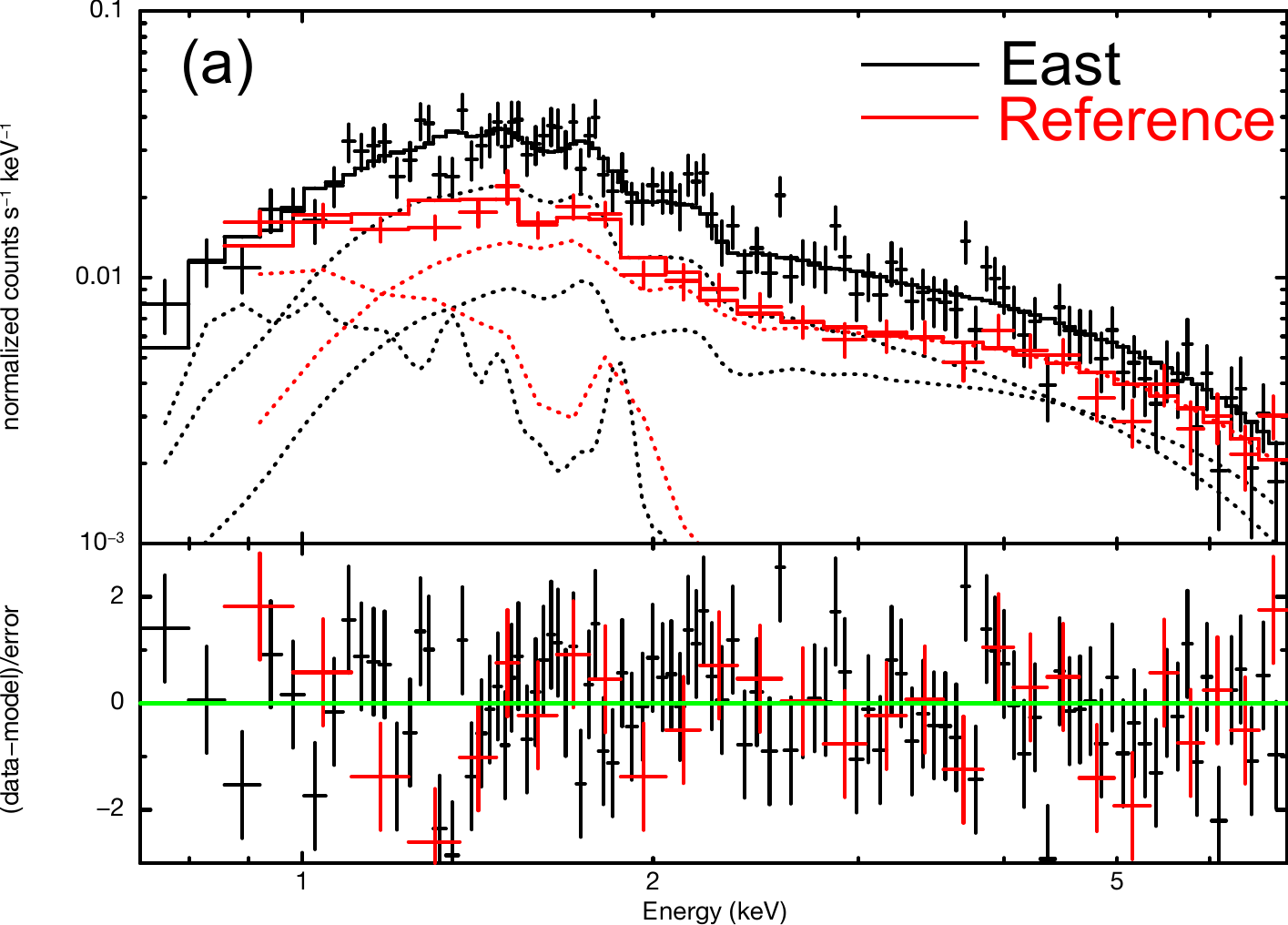} \plotone{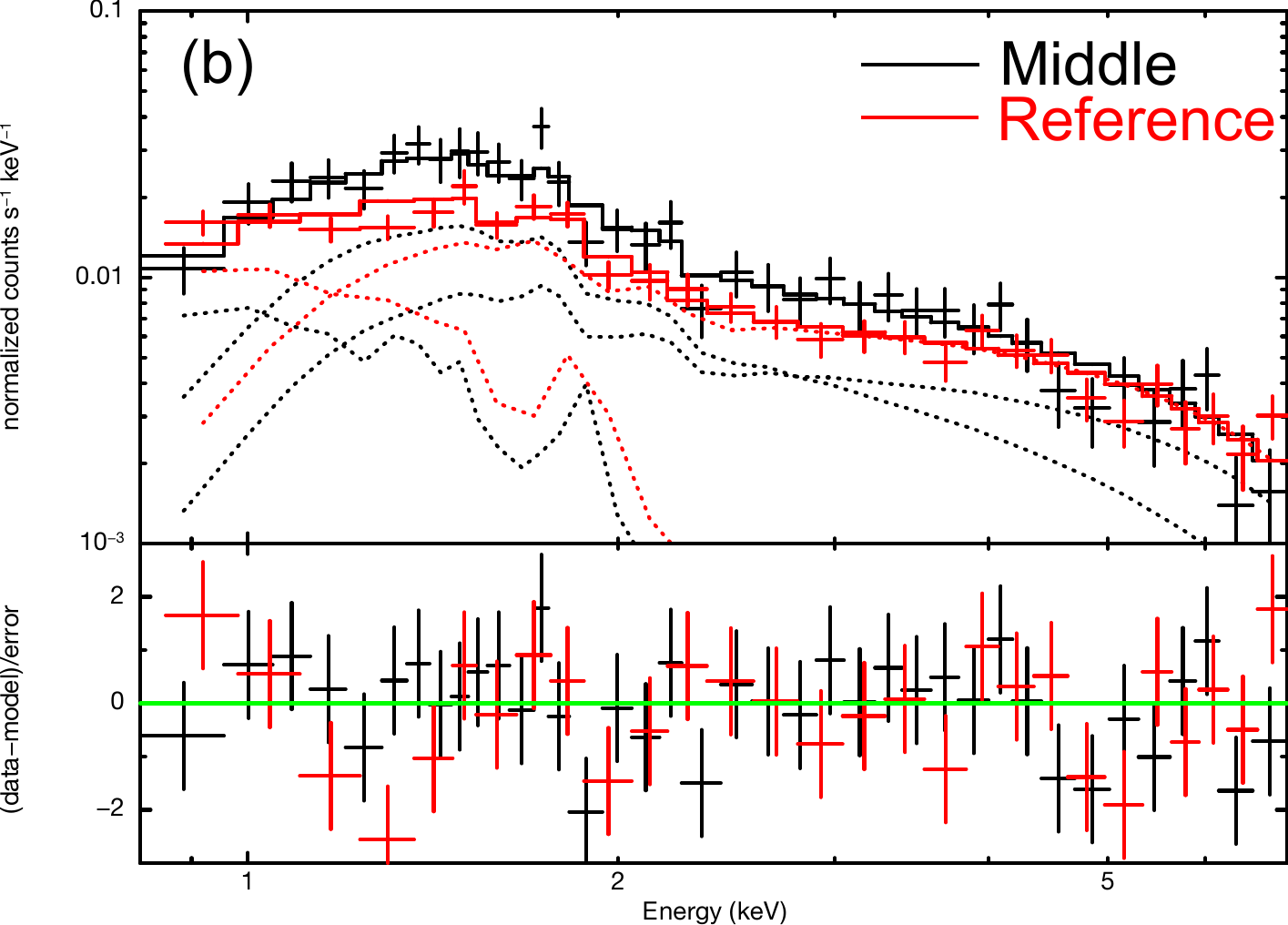} \plotone{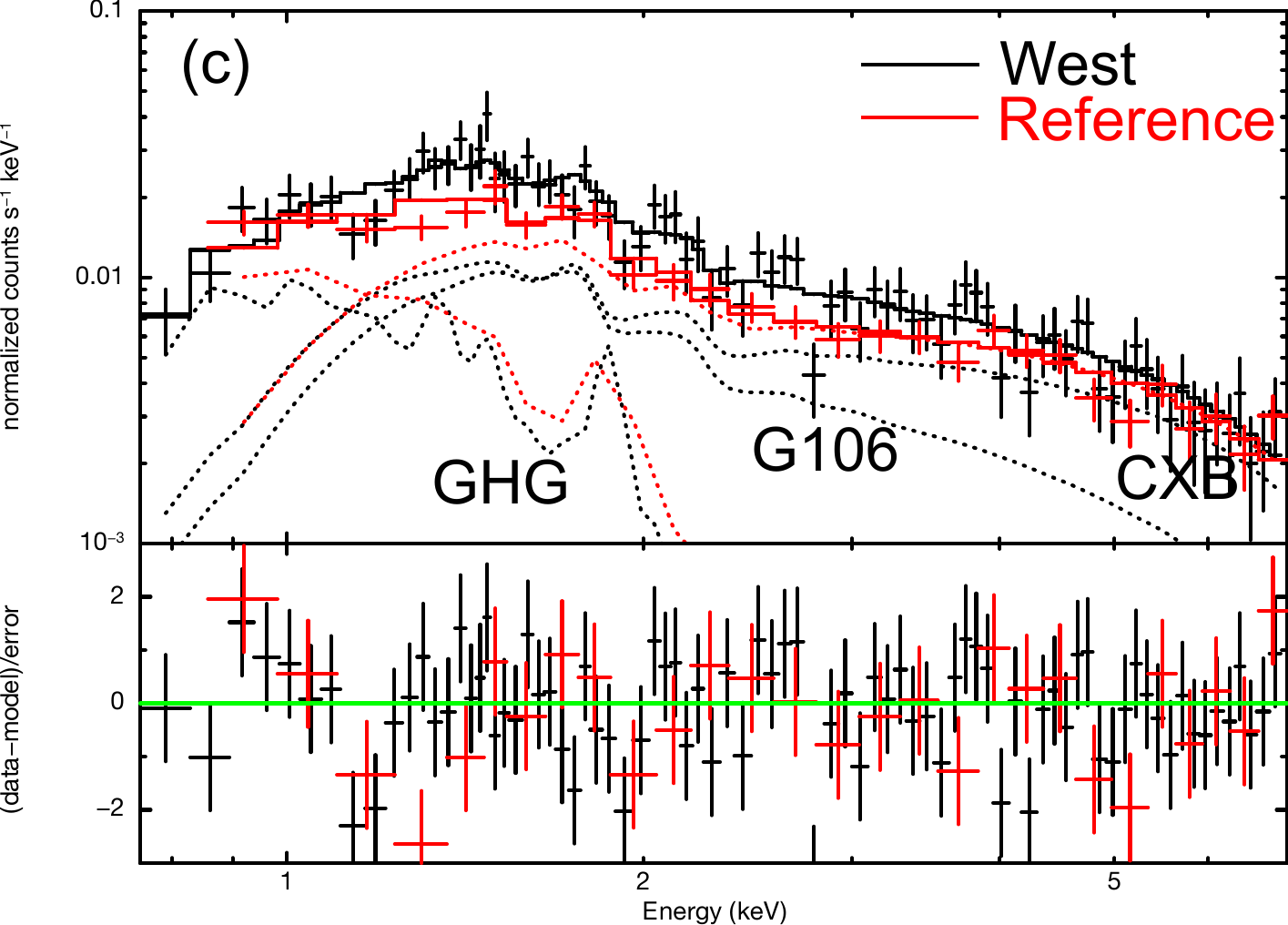}
\caption{XIS~3 spectra of the Source fields (black marks) fitted with
the nonthermal model (black solid lines, see
equation~(\ref{eq:specNTH})), and those of the Reference field (red
marks) fitted with the model omitting the emission from G106.3+2.7 (red
solid lines).  (a) East, (b) Middle, and (c) West fields. Contributions
from the GHG, CXB, and G106.3+2.7 (G106) are shown by dotted lines.}.
\label{fig:spec}
\end{figure}

We first study the spectra of individual fields. In
Figure~\ref{fig:spec}, we present the XIS spectra of the three Source
fields that are fitted with the nonthermal model
(equation~(\ref{eq:specNTH})). We display only XIS~3 spectra for
clarity. We also show the spectrum of the Reference field fitted with
the model omitting the emission from G106.3+2.7. Table~\ref{tab:fit}
shows the results of the spectral fitting. The temperature of the GHG is
represented by $kT_{\rm GHG}$ and it does not depend much on the
fields. We have confirmed that $kT_{\rm GHG}$ and the flux of the GHG
component are consistent with those obtained by the ROSAT All-Sky
Survey \citep{2019ascl.soft04001S}.
We have also compared our derived
normalizations of the CXB component (${\tt pegpwrlw}\langle 4\rangle$ in
equation~(\ref{eq:specNTH})) with the averaged value obtained by a
previous study ($6.38\times 10^{-8}\rm\: erg\: cm^{-2}\: s^{-1}\:
sr^{-1}$ for 2--10~keV; \citealt{2002PASJ...54..327K}) and found that
they are consistent with an error of a few percent.  The combined
absorption by the interstellar medium and possible dense gas around
G106.3+2.7 ($N_{\rm H,c}$) is close to that by the interstellar medium
alone ($N_{\rm H,i}=0.8\times 10^{22}\rm\: cm^{-2}$). This means that
the local absorption around G106.3+2.7 is ignorable. In some cases,
$N_{\rm H,c}$ is slightly smaller than $N_{\rm H,i}$, which may indicate
that $N_{\rm H,i}$ varies among fields. We also fit the XIS spectra with
the thermal model (equation~(\ref{eq:specTH})) and the results are given
in Table~\ref{tab:fit}.

Table~\ref{tab:fit} shows that the variations of $\Gamma$ (nonthermal)
and $kT$ and $Z$ (thermal) are small among the fields. The values of
$\chi^2$ in Table~\ref{tab:fit} indicate that the emission from
G106.3+2.7 can be represented by both nonthermal and thermal processes.
However, it is notable that $Z$ is much smaller than $1\: Z_\odot$
regardless of the fields. In order to obtain a stronger constraint, we
studied the combined spectrum of the three Source fields assuming that
their spectra are the same except for the normalizations. The result
shows that $Z=0.05_{-0.05}^{+0.08}\: Z_\odot$
(Table~\ref{tab:fit}). Considering that G106.3+2.7 is an object in the
Milky Way and also relatively close to the Sun, the abundance
is unrealistically small. Thus, we assume that the emission from
G106.3+2.7 is nonthermal from now on. The combined flux of the emission
({\tt pegpwrlw}$\langle 5\rangle$ in equation~(\ref{eq:specNTH})) from
the three fields is $3.6\pm 0.1\times 10^{-12}\rm\: erg\: cm^{-2}\:
s^{-1}$ in the 2--10~keV band\footnote{This value is larger than
$1.56\times 10^{-12}\rm\: erg\: cm^{-2}\: s^{-1}$ obtained by
\citet{2001ApJ...547..323H}. This is because \citet{2001ApJ...547..323H}
set a background region down to a radius of $5\farcm 25$. Our results
show that this region still contains diffuse emission from the
source. Thus, it is not appropriate to discuss the spectral energy
distribution (SED) of G106.3+2.7 based on the results of
\citet{2001ApJ...547..323H}.}.

\section{Discussion}
\label{sec:disc}

\subsection{CR energy spectrum}
\label{sec:spec}

In Table~\ref{tab:fit}, we present the average surface brightnesses
$\langle I\rangle$ of the source component (${\tt pegpwrlw}\langle
5\rangle$ in equation~(\ref{eq:specNTH}) or ${\tt apec}\langle 5\rangle$
in equation~(\ref{eq:specTH})) in the 2--10~keV band. The brightness
decreases from East to West, which means that the diffuse emission
becomes weaker as the distance from PSR~J2229+6114 increases.  We did
not study surface brightness distributions within each field because of
poor photon statistics. The trend is the same as that for the 1.4~GHz
radio continuum (\citealt{2000AJ....120.3218P, 2001ApJ...560..236K,
2003AJ....125.3145T}; see Figure~\ref{fig:map}), which suggests that the
diffuse X-ray emission originates from the pulsar as the radio emission
does, and that the density of CRs decreases as they disperse into the
interstellar space.  On the contrary, the gamma-ray emission is
coincident with a molecular cloud that is located around the Middle and
West fields (\citealt{2009ApJ...703L...6A,2019ApJ...885..162X}; see
Figure~\ref{fig:map}). These may indicate that the gamma-ray emission is
due to hadronic processes, while the X-ray and radio emissions are due
to leptonic ones.

Based on this idea, we study the spectral energy distribution (SED) of
G106.3+2.7. For the energy distribution of particles, we assume an
exponential cutoff power-law form:
\begin{equation}
 \frac{dN_i}{dE_i}\propto E^{-\alpha_i}
\exp\left(-\frac{E}{E_{{\rm cut},i}}\right)
\end{equation}
for both electrons ($i=e$) and protons ($i=p$). In this equation,
$\alpha_i$ is the spectral index, $E$ is the particle energy, and
$E_{{\rm cut},i}$ is the cutoff energy.  The parameters of spectral fit
are $\alpha_i$, $E_{{\rm cut},i}$, the total energy of particles above
1~GeV ($W_i$), the magnetic fields ($B$), and the target gas density
($n_{\rm gas}$). We note that there are many parameters for the fit and
some parameters are degenerated. Thus, we focus on the tendency
of the SED when the parameters are changed. We calculate radiative
processes for electrons using the models by \citet{2008MNRAS.384.1119F}
and we derive gamma-ray spectra using the models by
\citet{2006ApJ...647..692K}, \citet{2006PhRvD..74c4018K}, and
\citet{2008ApJ...674..278K}.

\begin{figure}[t]
\plotone{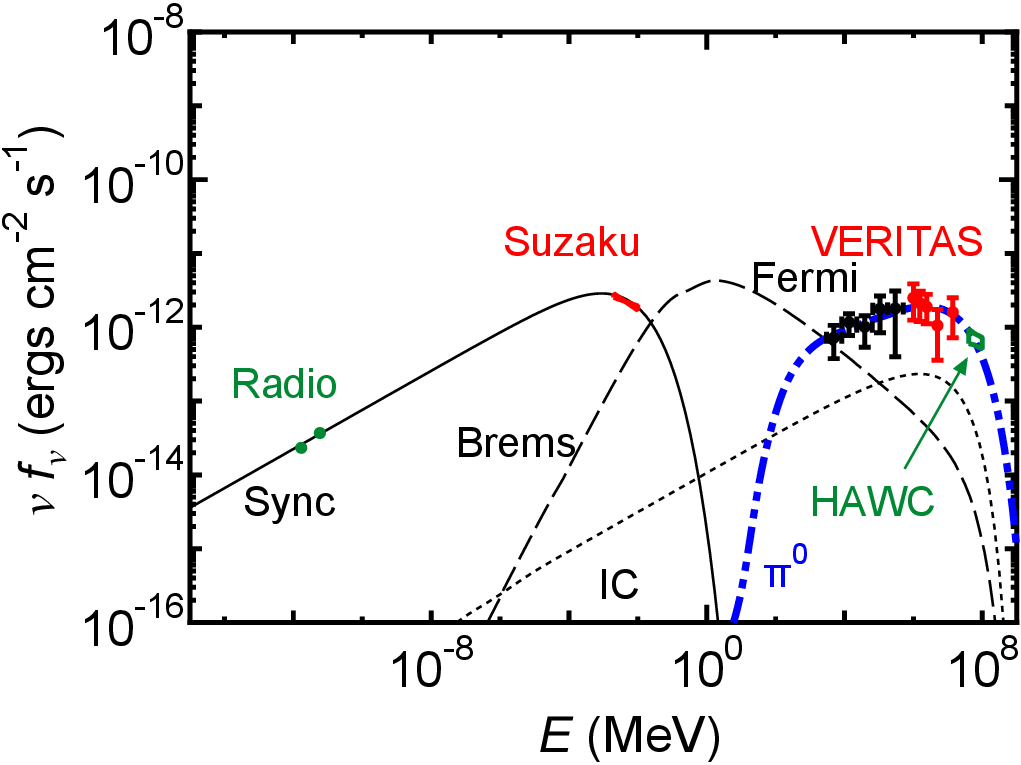} \caption{Spectral fit to the SED of G106.3+2.7 for the
hadronic-leptonic hybrid scenario. The thin black solid, dashed, and
dotted lines are for the synchrotron, nonthermal Bremsstrahlung, and IC
scattering, respectively. The thick blue two-dotted-dashed line is for the
$\pi^0$ decay gamma rays. Radio \citep{2000AJ....120.3218P}, 
Suzaku (this study), Fermi \citep{2019ApJ...885..162X}, VERITAS
\citep{2009ApJ...703L...6A} and HAWC \citep{2020ApJ...896L..29A}
observations are also shown. VERITAS results are scaled up by a factor
of 1.67 following \citet{2020ApJ...896L..29A}. \label{fig:HL}}
\end{figure}

Figure~~\ref{fig:HL} is one example of the fit. We assumed that the
distance to G106.3+2.7 is 800~pc \citep{2001ApJ...560..236K}.  Following
previous studies \citep[e.g.][]{2019ApJ...885..162X}, the average
density of background cloud gas is assumed to be $n_{\rm gas}=10\rm\:
cm^{-3}$ for all regions for the sake of simplicity.  In this case, the
density cannot be much larger than $n_{\rm gas}=10\rm\: cm^{-3}$ because
Bremsstrahlung emission from electrons becomes prominent in the gamma
rays and the shape of the spectrum is inconsistent with the gamma-ray
observations. However, we note that if the electrons exist in a lower
density region separately from the gamma-ray emitting protons, this
constraint is relaxed.  The indices $\alpha_e=2.5$ and $\alpha_p=1.9$
are well-constrained. The cutoff energies are $E_{{\rm cut},e}=80$~TeV
and $E_{{\rm cut},p}=400$~TeV. The former is broadly determined by our
X-ray observations. The magnetic field ($B=12\:\mu\rm G$) and the total
electron energy ($W_e=5.8\times 10^{47}\rm\: erg$) are degenerated by each
other for given synchrotron emission. However, $B$ cannot be much
smaller than this, because a smaller $B$ requires a larger $W_e$, which
leads to larger inverse Compton (IC) scattering from the electrons and
invalidates our assumption that the gamma-ray emission is of hadronic
origin. For the IC scattering process, an infrared (IR) radiation field
($T=30$~K and the energy density of $0.3\rm\: eV cm^{-3}$;
\citealt{2006ApJ...640L.155M}) is included in addition to the cosmic
microwave background (CMB). The total proton energy is $W_p=4.0\times
10^{47}(n_{\rm gas}/{\rm 10~cm^{-3}})^{-1}\rm\: erg$. We have confirmed
that the emissions from secondary electrons created through the
proton-proton interaction are ignorable.

\begin{figure}[t]
\plotone{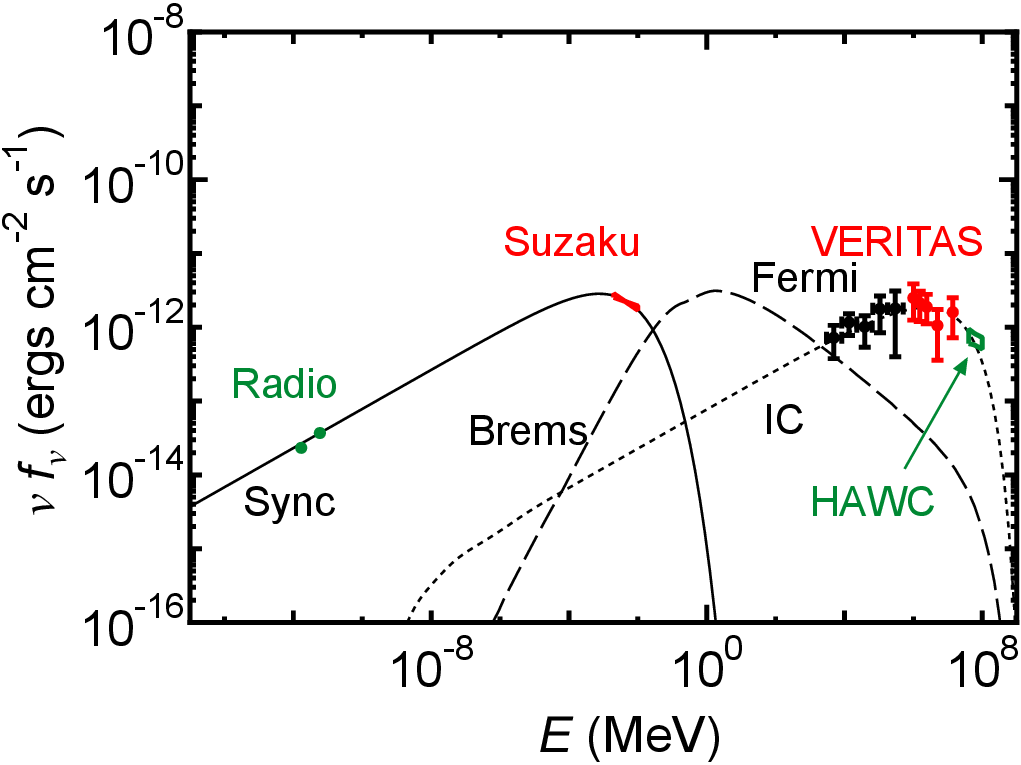} \caption{Same as Figure~\ref{fig:HL} but for the pure
leptonic scenario. \label{fig:LL}}
\end{figure}

We note that a pure leptonic model can also reproduce the observed SED
if we ignore the difference of spatial distributions between gamma rays
and other emissions (Figure~\ref{fig:LL}). Assuming that $n_{\rm
gas}=1\rm\: cm^{-3}$, we obtain $\alpha_e=2.5$, $E_{{\rm
cut},e}=130$~TeV, $W_e=4.2\times 10^{48}\rm\: erg$, and $B=3.9\:\mu\rm
G$. This means that it is difficult to reject pure leptonic models with
one-zone approaches based on the current observational data. 
However, the difference of the spatial distributions we found shows that
the source of the gamma rays is distinguished from that of the radio and
X-ray emissions. This supports the idea that the gamma ray is of
hadronic origin and that G106.3+2.7 is a PeVatron.

\subsection{Origin of the Nonthermal Emissions}

Our Suzaku observations show that the diffuse nonthermal X-ray emission
is bright around PSR~J2229+6114 (section~\ref{sec:spec}) and the photon
index $\Gamma$ is almost the same among the three fields
(Table~\ref{tab:fit}). These suggest that high-energy electrons ($E\sim
E_{{\rm cut},e}\sim 80$~TeV) are accelerated at the pulsar and they are
not affected by radiative cooling before they reach the West field. The
synchrotron cooling time of the electrons is given by
\begin{equation}
\label{eq:tsyn}
t_{\rm syn}  \sim 1100\: \left(\frac{E}{\rm 80~TeV}\right)^{-1}
\left(\frac{B}{12\: \mu\rm G}\right)^{-2}\rm\: yr~,
\end{equation}
and is much shorter than the characteristic age of PSR~J2229+6114
($t_c\sim 10$~kyr; \citealt{2001ApJ...552L.125H}). The short cooling
time suggests that the electrons need to travel fairly fast around the
pulsar.  This is the same for other old pulsars
\citep{2009PASJ...61S.189U,2010ApJ...719L.116B}.

We assume that G106.3+2.7 and PSR~J2229+6114 originate from the same
supernova explosion and that the electrons propagate through diffusion.
The diffusion coefficient of CR particles is given by
\begin{equation}
\label{eq:diff}
 D(E) = 10^{28}\chi \left(\frac{E}{\rm 10~GeV}\right)^{\delta}
\left(\frac{B}{3~\mu\rm G}\right)^{-\delta}\rm~ cm^2~s^{-1}~,
\end{equation}
where $\delta=1/2$ and $\chi (<1)$ is the reduction factor
\citep{2009ApJ...707L.179F,2009MNRAS.396.1629G}. We assume that
equation~(\ref{eq:diff}) can be applied to both CR electrons and protons
because the diffusion is probably caused by magnetic fluctuations and
the gyroradius of relativistic particles does not depend on their mass
explicitly.  The reduction factor $\chi$ may be associated with the
amplification of the magnetic fluctuations
\citep[e.g.][]{2010ApJ...712L.153F,2011MNRAS.415.3434F}. The
characteristic length scale that a diffusing particle of energy $E$
travels in time $t$ is given by $\ell_{\rm diff} = 2\sqrt{D(E) t}$. The
apparent distance between PSR~J2229+6114 and the West field is $\sim
8$~pc. Thus, for $E\sim E_{{\rm cut},e} \sim 80$~TeV and $B\sim
12~\mu\rm G$, the electrons can travel the distance without cooling if
$\chi \gtrsim 0.01$. The lower limit is comparable to the typical one
estimated around SNRs ($\chi\sim 0.01$;
\citealt{2009ApJ...707L.179F}). Similar values of $\chi$ could be
applied to other pulsars that are not affected by cooling
\citep{2009PASJ...61S.189U,2010ApJ...719L.116B}.

\citet{2020ApJ...896L..29A} suggested that high-energy protons that are
responsible for the gamma-ray emission might be accelerated by the SNR
G106.3+2.7.  Since the high-energy protons are likely to be accelerated
in the very early phase of SNR evolution
\citep[e.g.][]{2005A&A...429..755P}, they must have been confined around
the SNR to the present. Assuming that $E\sim E_{{\rm cut},p} \sim
400$~TeV, $B\sim 12~\mu\rm G$, and $t\sim t_c\sim 10$~kyr, the proton
diffusion length is $\ell_{\rm diff}\gtrsim 40$~pc for $\chi\gtrsim
0.01$ and it is much larger than the size of the gamma-ray emitting
region ($\sim 6$~pc). This means that the protons are not confined.
Thus, the SNR shock may not
be the origin of the protons unless the age is much smaller than
10~kyr. However, this problem may be solved if the gamma-ray emitting
cloud is three-dimensionally separated from the synchrotron emitting
region and the protons accelerated by the SNR have been trapped in the
cloud with an extremely small diffusion coefficient.

Another candidate of the accelerator of the CR protons is PSR~J2229+6114
or its PWN (\citealt{2019ApJ...885..162X}, see also
\citealt{1990JPhG...16.1115C,2008MNRAS.385.1105B,2020A&A...635A.138G}). If
this is the case, the protons may propagate from the pulsar in the
southwest direction in the same way as electrons. Some of them hit the
molecular cloud and create the observed gamma rays. However, even if
10\% of the spin-down luminosity of the pulsar ($L_{\rm sp} = 2.2\times
10^{37}\rm erg~s^{-1}$) is consumed to accelerate protons and all of the
protons enter the molecular cloud, the protons have to be confined for
$\sim 10$~kyr to explain the total proton energy in the molecular cloud
($W_p\sim 4.0\times 10^{47}\rm\: erg$). This cannot be realized if we
assume the diffusion coefficient discussed above. Thus, we may need to
consider some more complicated scenarios \citep[e.g.][]{2021arXiv210301814B}. For example, the interaction
between the PWN and the SNR may reaccelerate the CRs
\citep{2018MNRAS.478..926O}. Interestingly, another PeVatron candidate
HAWC~J1825-134 is also associated with two powerful pulsars
\citep{2020arXiv201215275A}.

We note that \citet{2006ApJ...638..225K} indicated that the
radio spectral index of the ‘‘Boomerang’’ PWN of PSR~J2229+6114
(Figure~\ref{fig:map}) is $\alpha=0.11\pm0.05$ ($S\propto
\nu^{-\alpha}$). The spectrum is much fatter than that of the overall
SNR represented by the tail ($\alpha\sim 0.6$--0.7;
\citealt{2000AJ....120.3218P}). The radio spectrum of the PWN becomes
steeper at larger radii, which may reflect synchrotron cooling
\citep{2006ApJ...638..225K}. In this case, this synchrotron component
should disappear outside the cooling radius that is about the size of
the PWN ($\sim 1\farcm 7$; \citealt{2001ApJ...547..323H}). Since the
radio tail extends much beyond the PWN, it seems to be a different
component from the PWN. The CR electrons that are associated with the
radio and X-ray emissions from the tail may have directly diffused out
of the pulsar. It is unlikely that they were accelerated by the SNR when
it was young considering the age ($\sim 10$~kyr) and the short cooling
time of the electrons (equation~(\ref{eq:tsyn})). Thus, another
possibility is that they are currently accelerated in situ, although
there is no clear shock-like structure in the middle-aged SNR
G106.3+2.7.

\section{Summary}
\label{sec:sum}

We have analyzed the Suzaku XIS data of the SNR G106.3+2.7. We
discovered diffuse X-ray emission and found that the brightness
increases toward the pulsar PSR~J2229+6114. While this trend is the same
as that of the radio emission, it is different from that of gamma-ray
emission concentrated on a molecular cloud. The X-ray emission can be
represented by either thermal or nonthermal processes. However, since
the extremely small metal abundance indicated by the thermal model is
unrealistic, the emission seems to be nonthermal synchrotron. We found
that the photon index of the nonthermal X-ray emission does not change
with the distance from the pulsar. This means that the parent electrons
are not affected by cooling and their diffusion is not very slow. The
parent protons of the gamma-ray emission may be tightly confined in the
cloud separately from the diffusing electrons.

\acknowledgments

We would like to thank the anonymous referee for very helpful
comments. This work was supported by MEXT KAKENHI No.18K03647, 20H00181
(Y.F.), 19K03908, 18H05459 (A.B.), 20K14491 (K.K.N.), 20H00175, 20H00178
(H.M.), and 20KK0071 (K.K.N. and H.M.). A.B. and K.K.N. were also
supported by Shiseido Female Researcher Science Grant and Yamada Science
Foundation, respectively.  The research presented in this paper has used
data from the Canadian Galactic Plane Survey, a Canadian project with
international partners, supported by the Natural Sciences and
Engineering Research Council.

\software{XSPEC \citep{1996ASPC..101...17A}, HEAsoft (v6.25;
\citealt{2014ascl.soft08004N}), sxrbg \citep{2019ascl.soft04001S}}

{\it Note added in Proof:} After submitting this draft, we
became aware of a preprint by \citet{2020arXiv201211531G} that studies
the same object. Although their Suzaku results are similar to ours,
their fluxes are somewhat different probably because of their different
background treatments.

\bibliography{G106_prep3}{}
\bibliographystyle{aasjournal}

\end{document}